\documentclass{PoS}

\usepackage{cite}

\newcommand{\be}{\begin{equation}}
\newcommand{\ee}{\end{equation}}
\newcommand{\ba}{\begin{eqnarray}}
\newcommand{\ea}{\end{eqnarray}}
\newcommand{\la}{\label}

\title{The leading hadronic contribution to (g-2) of the muon: The chiral behavior using the mixed representation method}

\ShortTitle{The chiral behavior of $a_\mu^{HLO}$ using the mixed representation method}

\author{\speaker{Anthony Francis}$^3$, Vera G\"ulpers$^3$, Gregorio Herdoiza$^{2,4}$, Hanno Horch$^2$, Benjamin J\"ager$^5$, Harvey~B.~Meyer$^{1,2,3}$ and Hartmut Wittig$^{1,2,3}$\\
        PRISMA Cluster of Excellence$^1$,
        Institut f\"ur Kernphysik$^2$ and Helmholtz~Institut~Mainz$^3$,
        Johannes Gutenberg-Universit\"at Mainz,
        D-55099 Mainz, Germany;\\
        Departamento de F\'isica Te\'orica and Instituto de F\'isica Te\'orica
    UAM/CSIC$^4$,\\ Universidad Aut\'onoma de Madrid, Cantoblanco, E-28049
    Madrid, Spain;\\
        Department of Physics, College of Science, Swansea University$^5$, SA2 8PP Swansea, UK\\
        E-mail: \email{francis@kph.uni-mainz.de},
        \email{guelpers@kph.uni-mainz.de},
        \email{gregorio.herdoiza@uam.es},
        \email{horch@kph.uni-mainz.de},
         \email{B.Jaeger@swansea.ac.uk},
        \email{meyerh@kph.uni-mainz.de},
        \email{wittig@kph.uni-mainz.de}}

\abstract{We extend our analysis of the leading hadronic contribution to the anomalous magnetic moment of the muon using the mixed representation method to study its chiral behavior. We present results derived from local-conserved two-point lattice vector correlation functions, computed on a subset of light two-flavor ensembles made available to us through the CLS effort with pion masses as low as 190\,MeV. The data is analyzed also using the more standard four-momentum method. Both methods are systematically compared as the calculations approach the physical point.
}

\FullConference{The 32nd International Symposium on Lattice Field Theory\\
                 23-28 June, 2014\\
                 Columbia University New York, NY}

\begin{document}

\section{Introduction}
The hadronic vacuum polarization $\Pi(Q^2)$ is of great importance in precision tests of the Standard Model of particle physics. It enters, for instance, the running of the QED coupling constant. Additionally, it currently represents one of the dominant uncertainties in the Standard Model prediction of the anomalous magnetic moment of the muon $(g-2)_\mu$.\\
The leading order hadronic contribution $a_\mu^{HLO}$ is accessible by computing the hadronic vacuum polarization function $\Pi(Q^2)$ and convoluting it with the an electro-weak kernel $K_E(Q^2,m_\mu)$ \cite{Jegerlehner:2009ry,gen0,gen1},
\begin{equation}\label{eq:gminus}
a_\mu^{HLO} = 4\alpha^2\int dQ^2 K_E(Q^2,m_\mu) \Big( \Pi(Q^2) - \Pi(0) \Big)~~.
\end{equation}
The leading hadronic contribution to the vacuum polarization $e^2\Pi(Q^2)$ in the spacelike domain can be expressed through the vector meson spectral function via a once-subtracted dispersion relation,
\begin{equation}\label{eq:disp}
\Big( \Pi(Q^2) - \Pi(0) \Big) = Q^2 ~\int_0^\infty~ds \frac{\rho(s)}{s(s+Q^2)} ~~.
\end{equation}
In the dispersive approach \cite{Jegerlehner:2009ry} one replaces the spectral function with the experimentally accessible $R(s)$-ratio by making use of the optical theorem.  
On the lattice the problem of determining $\widehat{\Pi}(Q^2)=4\pi^2(\Pi(Q^2) - \Pi(0))$ can be approached both from the left and the right hand side of Eq.~\ref{eq:disp}.\\
In the following, we extend our analysis of the leading hadronic contribution to the anomalous magnetic moment of the muon using both the four-momentum method \cite{gen1, gen2, gen3, gen4, gen5, gen6, DellaMorte:2012cf} (i.e. by evaluating the lhs of Eq.~\ref{eq:disp}) and the recently introduced mixed representation method \cite{gen9,Francis:2013fzp,Bernecker:2011gh} (i.e. by evaluating the rhs of Eq.~\ref{eq:disp}). 
We systematically compare both methods and monitor their approach towards the physical point. The preliminary results presented here are derived from local-conserved two-point lattice vector correlation functions, computed on a subset of light two-flavor ensembles made available to us through the CLS effort with pion masses as low as 190\,MeV.


\section{The four-momentum method to compute $\widehat{\Pi}(Q^2)$}
On a Euclidean lattice the vacuum polarization tensor can be defined as the four dimensional Fourier transform of the vector current-current correlation function:
\be\la{eqn:fourcorr}
\Pi_{\mu\nu}(Q) \equiv \int d^4x \, e^{iQ\cdot x} \langle j_\mu(x) j_\nu(0)\rangle.
\ee
Here $O(4)$ invariance and current conservation imply the tensor structure
\be
\la{eqn:tensorstruct}
\Pi_{\mu\nu}(Q) =\big(Q_\mu Q_\nu -\delta_{\mu\nu}Q^2\big) \Pi(Q^2).
\ee
We can therefore extract $\Pi(Q^2 \geq Q^2_{latt,min}(L))$ from lattice calculations.
However, computing $\widehat{\Pi}(Q^2)$ following this recipe, one is faced with the problem that the additive renormalization $\Pi(Q^2=0)$ is not directly available since the lowest available $Q^2_{min}=Q^2_{latt,min}(L)$ is dictated by the lattice discretization.
Consequently it has to be estimated using an extrapolation procedure $Q^2_{min} \rightarrow Q^2=0$, using e.g. a Pad\'e Ansatz \cite{gen6,DellaMorte:2012cf,gen7,gen8}. In addition the integrand of Eq.~\ref{eq:gminus} is strongly peaked around the mass of the lepton, and with the muon mass at $m_\mu\simeq105.65$\,MeV \cite{PDG}, this is generally below the lattice momentum resolution. As a consequence the resulting value of $a_\mu^{HLO}$ depends crucially on the correct extrapolation and therefore precision lattice data at low momentum $Q^2$ \cite{gen7, gen8}.

\section{The mixed representation method to compute $\widehat{\Pi}(Q^2)$}
In addition to the four-momentum method, in \cite{gen9,Francis:2013fzp,Bernecker:2011gh} a new method to compute $\widehat\Pi(Q^2)$ without the problem of having to estimate $\Pi(Q^2=0)$ was proposed. It has the advantage that it can be used to calculate any value of the virtuality $Q^2$. To this extent one approaches $\widehat{\Pi}(Q^2)$ from the right hand side of Eq.~\ref{eq:disp} by noting the electromagnetic spectral function $\rho(s)$ is directly linked to the lattice vector meson current-current correlator $\langle j_\mu(x) j_\nu(0)\rangle$ in the mixed time-momentum representation
\begin{equation}
G(x_0, \vec k) \stackrel{\mu=\nu}{=} \int d^3x\,e^{i\vec k \vec x} \langle J_\mu(x_0,\vec x) J_\nu(0) \rangle = \frac{1}{2} \int_0^\infty ds \sqrt{s}\rho(s)  e^{-\sqrt{s}|x_0|}~~.
\end{equation}
Exploiting this observation one arrives at an expression for $\widehat{\Pi}(Q^2)$ in terms of an integral over Euclidean time of the mixed representation correlator \cite{Francis:2013fzp}:
\begin{equation}\label{eq:mixrep}
4\pi^2\Big( \Pi(Q^2) - \Pi(0) \Big) = 4\pi^2\,\int_0^{\infty} dx_0\, G(x_0,\vec k=0)\, \Big[ x_0^2 - \frac{4}{Q^2}\sin^2(\frac{1}{2}Qx_0)\Big]~~.
\end{equation}
Additionally $a_\mu^{HLO}$ can be evaluated directly without having to take the intermediate step of calculating $\widehat{\Pi}(Q^2)$ by plugging Eq.~\ref{eq:mixrep} into Eq.~\ref{eq:gminus}, see \cite{Francis:2013fzp,Bernecker:2011gh}. However, for comparing both methods in the following, we find the extra step of computing $\widehat{\Pi}(Q^2)$ useful.\\
For a rigorous result, Eq.~\ref{eq:mixrep} must be integrated for all Euclidean times $t\rightarrow \infty$, a requirement that cannot be fulfilled on a finite lattice.
However, since the correlator drops exponentially with time, the integral can be truncated with only a small cost in accuracy, provided that the lattice data is precise enough for a large enough time separation. Here, we aim to estimate the large distance part of the integral in Eq.~\ref{eq:mixrep} by extrapolating the vector correlator to its asymptotic behavior.
Consequently the result for $a_\mu^{HLO}$ depends crucially on the knowledge of the long distance correlator or equivalently the low lying spectrum. In principle, the difficulty of extrapolating $Q^2_{min} \rightarrow Q^2=0$ in the four-momentum method has been replaced by precisely determining the large distance behavior of the mixed representation correlator.

\section{Numerical Setup}

In the following, we study the chiral behavior of  $a_\mu^{HLO}$ and $\widehat{\Pi}(Q^2)$ using both methods on dynamical gauge
configurations with two mass-degenerate quark flavors.  The gauge
action is the standard Wilson plaquette action \cite{Wilson:1974sk},
while the fermions were implemented via the O($a$) improved Wilson
discretization with non-perturbatively determined clover coefficient
$c_{\rm sw}$ \cite{Jansen:1998mx}.  The configurations
were generated within the CLS effort using algorithms
based on the DD-HMC\cite{CLScode} and MP-HMC packages\cite{mphmc}.  
We calculated local-conserved
correlation functions using the same discretization and masses as in
the sea sector on a set of lattice ensembles with $\beta=5.30$ entailing a lattice spacing of $a =
0.0631(21)$\,fm~\cite{Capitani:2011fg} and pion masses ranging between $m_\pi=451$\,MeV down to $m_\pi=190$\,MeV, see Tab.~\ref{tab:par} for a list of lattice parameters.
Note, in the following, all correlation functions for strange quark masses are available as partially quenched, valence observables.

\begin{table}[h!]
\begin{center}
\begin{tabular}{cccccc}
    	\hline
    	Lattice size & $L$ $[\mathrm{fm}]$ & 
    	$m_\pi$ $[\mathrm{MeV}]$ & $m_\pi L$ & $N_{meas}(N_{conf})$  & Label\\
    	\hline
    	$64 \times 32^3$ & $2.0$ & $451$ & $4.7$ & $4000(1000)$ &  E5 \\
    	$96 \times 48^3$ & $3.0$ & $324$ &  $5.0$ & $1200(300)$ &  F6 \\
    	$96 \times 48^3$ & $3.0$ & $277$ &  $4.2$ & $1000(250)$ &  F7 \\
    	$128 \times 64^3$ & $4.0$ & $190$ & $4.0$ & $820(205)$ &  G8  \\
    	\hline
\end{tabular}
\caption{Table of lattice parameters. To study the chiral behavior the set of ensembles used is fixed at $\beta=5.30$ and a lattice spacing of $a =
0.0631(21)$\,fm~\cite{Capitani:2011fg}. All correlators were calculated with four sources per configuration.}
\label{tab:par}
\end{center}
\end{table}

\section{Numerical Results}

\begin{figure}[h!]
\centering
\includegraphics[width=0.45\textwidth]{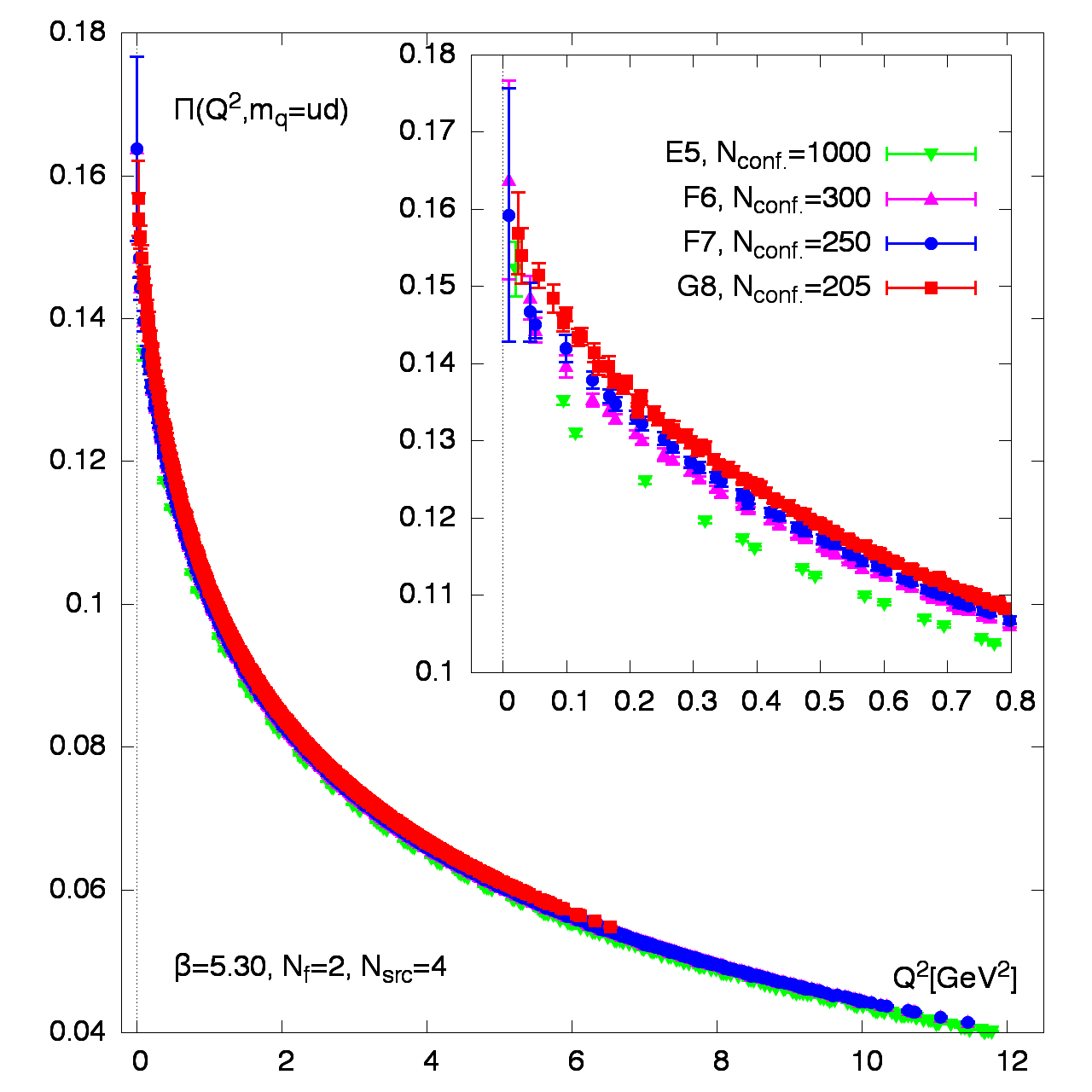}
\includegraphics[width=0.45\textwidth]{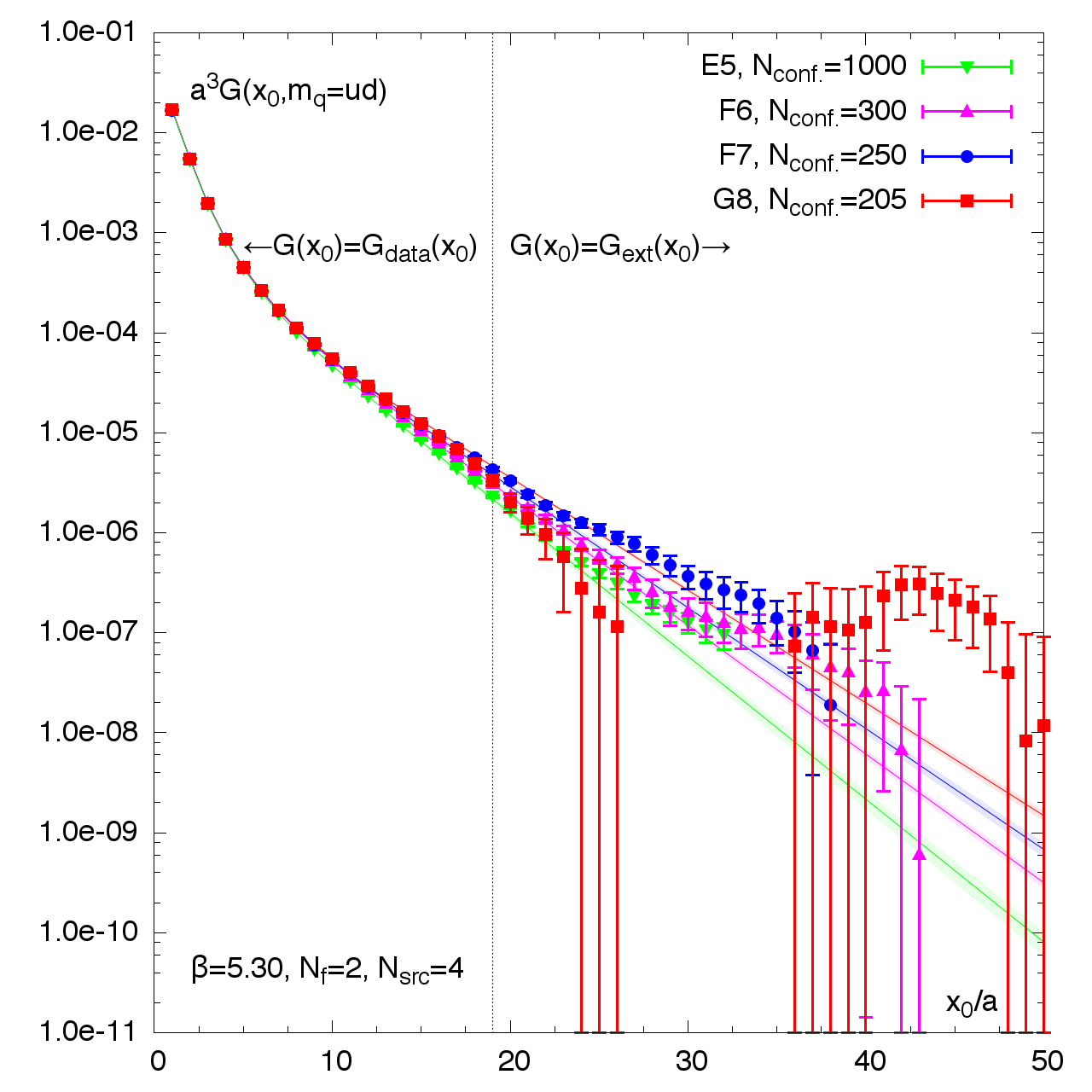}
\caption{ Left: Updated results for $\Pi(Q^2)$ using the four-momentum method with $a =
0.0631(21)$fm and pion masses ranging between $m_\pi=451$\,MeV and $m_\pi=190$\,MeV. The insert shows the especially interesting low $Q^2$ region. Right: The mixed representation vector meson correlator on the same lattice ensembles. At $x_0/a=19$ we extend the lattice data by a single exponential, since the signal is lost beyond this distance at our current level of statistics.}
\label{fig:update}
\end{figure}

Updating our ongoing programme to compute $a_\mu^{HLO}$, in Fig.~\ref{fig:update} (left) we show the results of $\Pi(Q^2)$ obtained using the four-momentum method. The calculation closely follows the procedure presented in \cite{gen6,DellaMorte:2012cf}, as such local-conserved currents were used to compute $\Pi_{\mu\nu}(Q)$. Following \cite{DellaMorte:2012cf,Bedaque:2004bx}, twisted boundary conditions were used to increase the set of available lattice momenta by three twist angles. In addition propagators were computed on four maximally separated source positions on every configuration. The small errors of the results in Fig.~\ref{fig:update} (left) with pion masses from $m_\pi=451$\,MeV down to $m_\pi=190$\,MeV indicate that we are capable of reaching very high precision for all but the lowest $Q^2$ values in this way. Turning to Fig.~\ref{fig:update} (right), where we show the corresponding mixed representation correlation functions, this decrease in accuracy manifests itself by a rapidly deteriorating signal in $G(x_0)$ for $x_0/a\geq 19$ or $x_0\geq 1.1$\,fm. This is linked to a substantial contribution in the low $Q^2$ region from the low lying spectrum in the vector channel, including the two-pion state. Therefore, the large distance behavior of $G(x_0)$, the sharp increase of errors in the low $Q^2$ region, are a reflection of the lattice data not being accurate enough to capture exactly this part of the spectrum. As a consequence, we conclude, that model-independent, precision results for $\Pi(Q^2)$ require improvement in the lattice determination of the low lying vector spectrum. In the mixed representation method one possibility is to concentrate on the low lying masses of the correlation function by setting up a GEVP with additional interpolating operators.\\
Based on the phenomenological observation, that one expects large contributions to $a_\mu^{HLO}$ up to distances of $\sim 1.5$\,fm \cite{Bernecker:2011gh, Francis:2013fzp}, the results in Fig.~\ref{fig:update} (right) indicate truncating Eq.~\ref{eq:mixrep} at $x_0\simeq 1.1$\,fm, will not suffice to obtain a full result of $a_\mu^{HLO}$. For this reason we smoothly extend the correlation function at $x_0\simeq 1.1$fm by assuming the ground state is reached at this point and decays with a single exponential. Although more elaborate models can be used to extend the correlation function, our present data does not warrant the introduction of additional assumptions.
In the following we will use the extended correlator and updated $\Pi(Q^2)$ to compute $a_\mu^{HLO}$ and $\widehat{\Pi}(Q^2)$. Note that the determination of the long-distance behavior of the vector
correlator involves a fit before this step. The four-momentum method, on the
other hand, relies on fitting the $Q^2$ behavior close to $Q^2=0$.

\section{Comparing the four-momentum and mixed representation methods}

\begin{figure}[t!]
\centering
\includegraphics[width=0.45\textwidth]{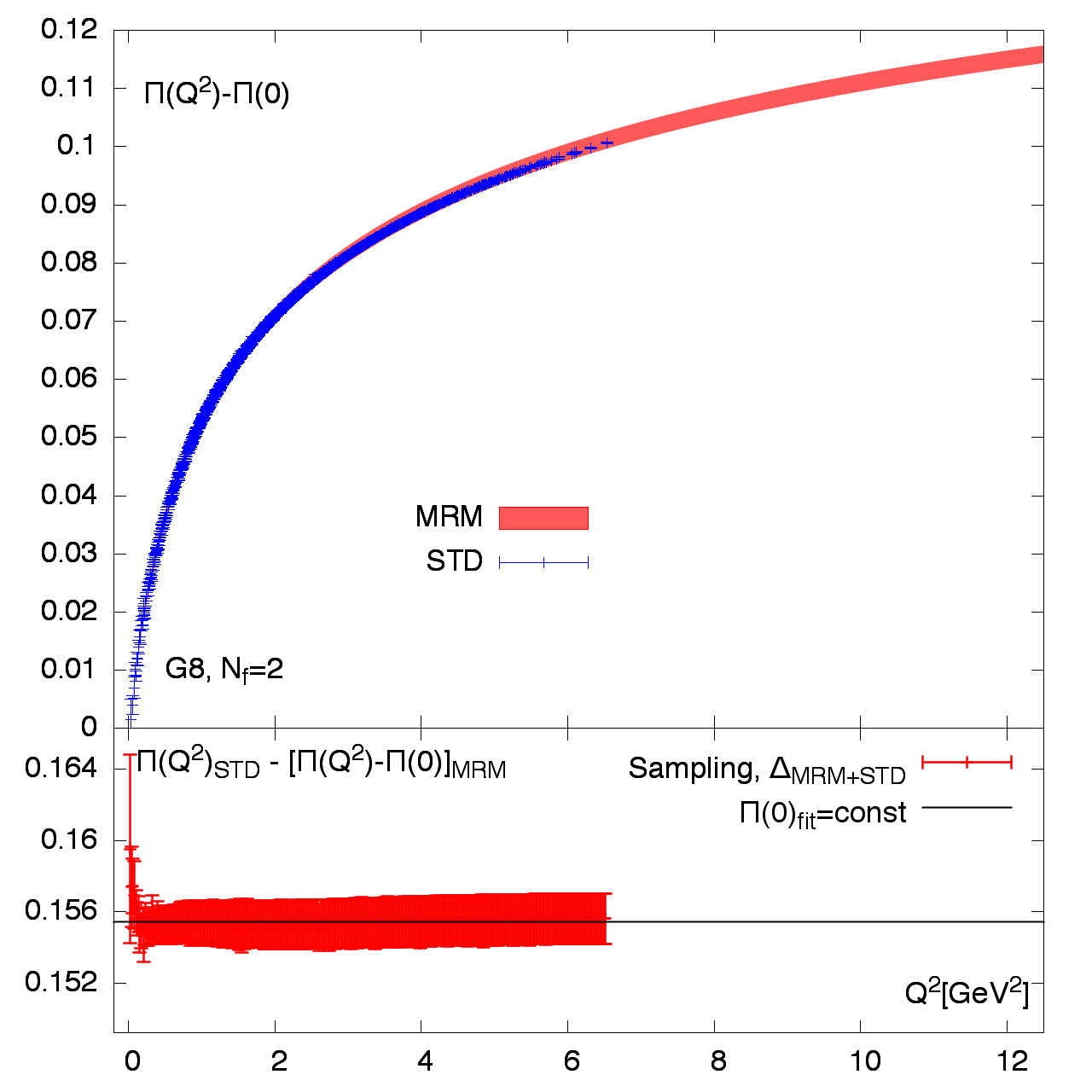}
\includegraphics[width=0.45\textwidth]{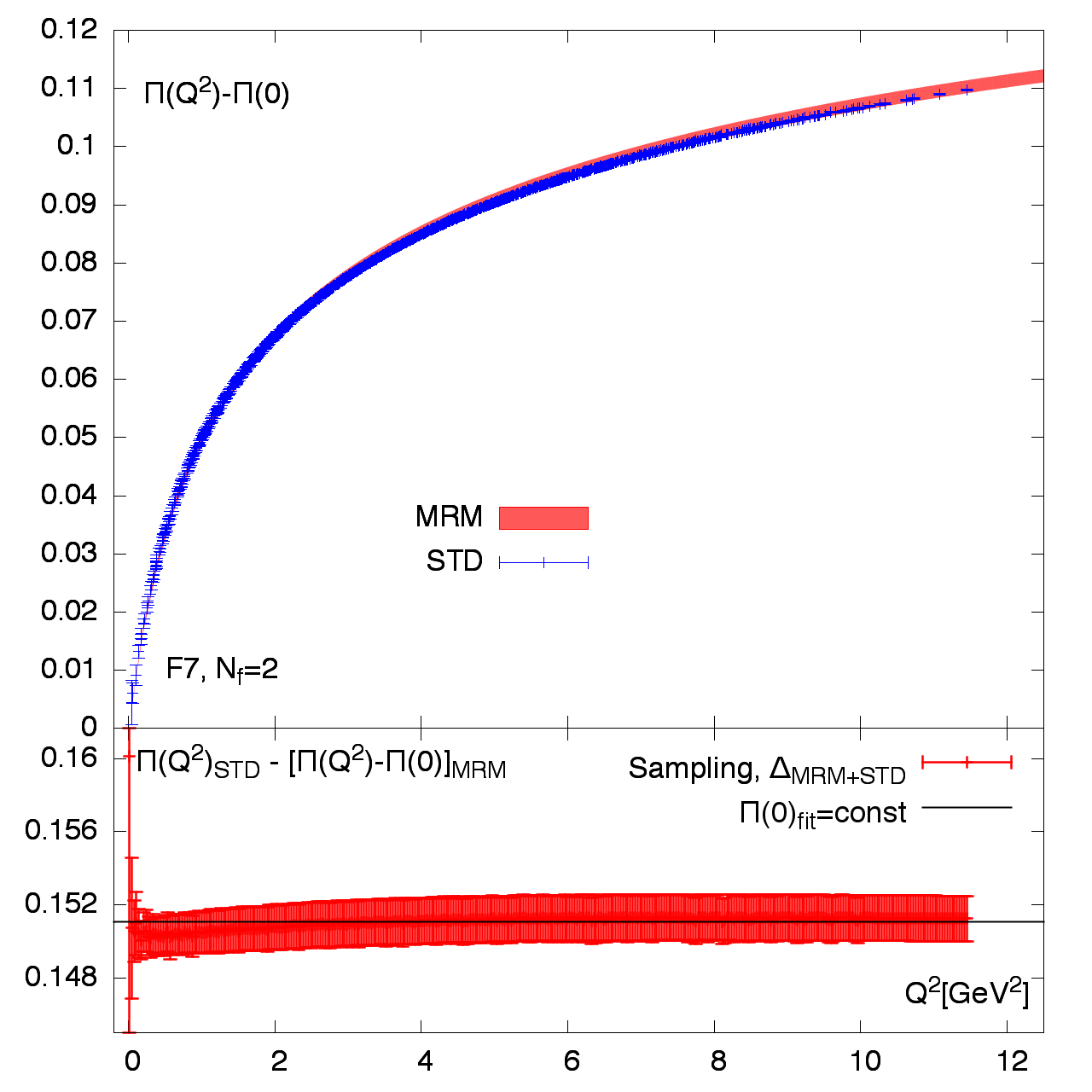}
\caption{ Calculating the difference of the HVP obtained from the four-momentum method and $\widehat{\Pi}(Q^2)$ from the mixed representation method, we arrive at a measure to monitor the systematics in the two different analyses. The top panels show $\widehat{\Pi}(Q^2)$ obtained from the four-momentum (STD) and mixed representation (MRM) methods on the G8 (left) and F7 (right) lattice ensembles. The bottom shows the difference, i.e. $\Pi(0)$, for the $Q^2$ available in the four-momentum method.}
\label{fig:combine}
\end{figure}

\begin{figure}[t!]
\centering
\includegraphics[width=0.6\textwidth]{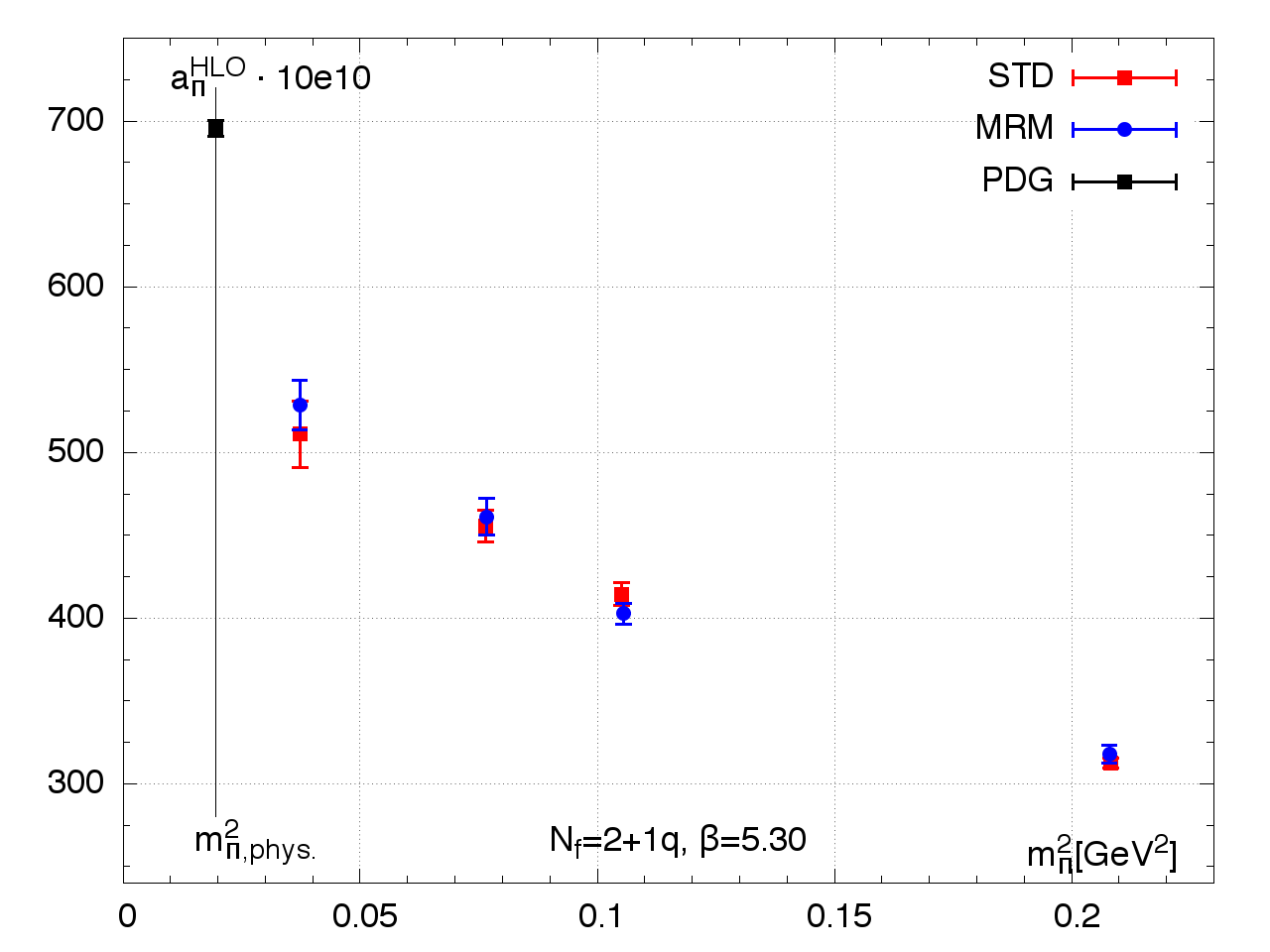}
\caption{The chiral behavior of $a_\mu^{HLO}$ in $N_f=2+1_q$ QCD at $a=0.0631(21)$\,fm. The shown results are obtained using the four-momentum (STD) and mixed representation methods (MRM). Together they give an estimate of $a_\mu^{HLO}$ taking into account the different analysis systematics. The physical point and PDG \cite{PDG} values are given in black for reference.}
\label{fig:amu}
\end{figure}

The four-momentum and the mixed representation methods are based on different treatments of lattice data for the vector
correlator. If both methods are controlled, they should yield consistent
results for $a_\mu^{HLO}$ and $\widehat{\Pi}(Q^2)$. Any deviation should
arise from the different systematics of the analysis machinery only. Hence, by imposing
$\widehat{\Pi}_{STD}(Q^2)=\widehat{\Pi}_{MRM}(Q^2)$ we can compute the
additive renormalization $\Pi(0)$, which can then be compared to the result
from extrapolating $\widehat{\Pi}_{STD}$ to $Q^2=0$,
\be\label{eq:pidiff}
\widehat{\Pi}_{STD}(Q^2) = \widehat{\Pi}_{MRM}(Q^2)
~~~\Rightarrow~~~ \Pi(0) = \Pi_{STD}(Q^2) -  \Big( \Pi(Q^2) - \Pi(0) \Big)_{MRM} ~~.
\ee
As a result are able to monitor the systematics due to the different analyses, while at the same time determining $\Pi(0)$. 
In Fig.~\ref{fig:combine} we show the results for $\widehat{\Pi}(Q^2)$ and $\Pi(0)$ obtained via Eq.~\ref{eq:pidiff} on the G8 (left) and F7 (right), i.e. $m_\pi=190$\,MeV and $m_\pi=277$\,MeV, lattice ensembles. 
To determine $\widehat{\Pi}(Q^2)$ in the four-momentum method we use a Pad\'e fit and follow the procedure outlined in \cite{gen6,DellaMorte:2012cf}.\\
The top panels of Fig.~\ref{fig:combine} show $\widehat{\Pi}(Q^2)$ obtained from the four-momentum (STD) and mixed representation (MRM) methods. Since $Q^2$ is discrete in the four-momentum method, these results are shown as points, while those obtained using the mixed representation method are given as bands. The bottom panels show the result of Eq.~\ref{eq:pidiff} over $Q^2$ for those $Q^2=Q^2_{STD}$ dictated by the four-momentum results. 
In the direct comparison of $\widehat{\Pi}(Q^2)$ the two methods show very good agreement. However, we find the results of Eq.~\ref{eq:pidiff} highlight the differences especially in the low $Q^2$ region, as the estimate of $\Pi(0)$ shows a decreasing trend around $Q^2\simeq 1\mathrm{GeV}^2$ with a sharp increase around the lowest available $Q^2$. Above $Q^2\simeq 2\mathrm{GeV}^2$ for G8 and $Q^2\simeq 4\mathrm{GeV}^2$ for F7 we furthermore observe a flat behavior of the results.
Since all deviations are within their respective errors, our results indicate a very good agreement of both methods, with systematics within the quoted errors.\\
Integrating our respective results for $\widehat{\Pi}(Q^2)$ via Eq.~\ref{eq:gminus} we are able to map out $a_\mu^{HLO}$ as it approaches the chiral limit using both methods.  
This is shown in Fig.~\ref{fig:amu}, whereby we added the quenched strange quark contribution to the quoted numbers. Taking the results of both methods together we can estimate $a_\mu^{HLO}$ taking into account the different analysis systematics.
  
\section{Conclusions}

We presented results on the leading order hadronic contribution to the anomalous magnetic moment of the muon as it approaches the chiral limit using two different analysis methods in lattice QCD. Comparing the four-momentum and mixed representation methods we find they serve as independent cross-checks of each other, since they process equivalent data. As a consequence they serve as a means to estimate the systematic uncertainties inherent in any calculation
performed in finite volume. We highlighted that a precision result of $a_\mu^{HLO}$ requires accurate knowledge of the asymptotic behavior of the vector meson current-current correlation function and discussed how this can be systematically achieved in the mixed representation method. In the future this will enable a precision determination of $a_\mu^{HLO}$ and will allow for a straightforward inclusion of the disconnected contributions.

\acknowledgments
We are grateful to our colleagues within CLS for sharing the lattice ensembles used. We thank Dalibor Djukanovic and Christian Seiwerth for their technical support. This work was granted access to the HPC resources of the Gauss Center for Supercomputing at
Forschungzentrum J\"ulich, Germany, made available within the Distributed European Computing
Initiative by the PRACE-2IP, receiving funding from the European Community's Seventh Framework
Programme (FP7/2007-2013) under grant agreement RI-283493 (project PRA039).
The correlation functions were computed 
on the dedicated QCD platforms ``Wilson''  at the Institute for Nuclear Physics,
University of Mainz, and ``Clover''  at the Helmholtz-Institut Mainz.
This work was supported by the \emph{Center for Computational Sciences}
as part of the Rhineland-Palatinate Research Initiative.


\begin{thebibliography}{99}
\setlength{\itemsep}{-0.1mm}

\bibitem{Jegerlehner:2009ry} F.~Jegerlehner and A.~Nyffeler, Phys.\ Rept.\  {\bf 477} (2009) 1.

\bibitem{gen0} E. de Rafael, Phys.\ Lett.\ B322 (1994) 239.
\bibitem{gen1} T. Blum, Phys.\ Rev.\ Lett.\ 91 (2003) 052001.
\bibitem{gen2} M. G\"ockeler et al. (QCDSF Collaboration), Nucl.\ Phys.\  B688 (2004) 135.
\bibitem{gen3} C. Aubin and T. Blum, Phys.\ Rev.\ D75 (2007) 114502.
\bibitem{gen4} X. Feng, K. Jansen, M. Petschlies, and D. B. Renner, Phys.\ Rev.\ Lett.\ 107 (2011) 081802.
\bibitem{gen5} P. Boyle, L. Del Debbio, E. Kerrane, and J. Zanotti, Phys.\ Rev.\ D85 (2012) 074504.
\bibitem{gen6} M. Della Morte, B. J\"ager, A. J\"uttner, and H. Wittig, JHEP {\bf 1203} (2012) 055.
\bibitem{DellaMorte:2012cf} M.~Della Morte, B.~Jager, A.~Juttner and H.~Wittig, PoS LATTICE {\bf 2012}, 175 (2012).

\bibitem{gen9} X. Feng, S. Hashimoto, G. Hotzel, K. Jansen, M. Petschlies, et al., Phys.\ Rev.\ D {\bf 88}, 034505 (2013).
 \bibitem{Bernecker:2011gh} D.~Bernecker and H.~B.~Meyer, Eur.\ Phys.\ J.\ A {\bf 47} (2011) 148.
 \bibitem{Francis:2013fzp} A.~Francis, B.~J\"ager, H.~B.~Meyer and H.~Wittig, Phys.\ Rev.\ D {\bf 88} (2013) 054502
\bibitem{PDG} J. Beringer et al. (Particle Data Group), Phys.\ Rev.\ D86 (2012) 010001.
\bibitem{gen7} G. de Divitiis, R. Petronzio, and N. Tantalo, Phys.\ Lett.\ B718 (2012) 589.
\bibitem{gen8} C. Aubin, T. Blum, M. Golterman, and S. Peris, Phys.\ Rev.\ D86 (2012) 054509.
 
\bibitem{Wilson:1974sk} K.~G.~Wilson, Phys.\ Rev.\ D10\ (1974) 2445.
\bibitem{Jansen:1998mx} K.~Jansen and R.~Sommer, Nucl.\ Phys.\ B530\ (1998) 185.
\bibitem{CLScode} http://luscher.web.cern.ch/luscher/DD-HMC/index.html
\bibitem{mphmc} M.~Marinkovic and S.~Schaefer, PoS\ LATTICE2010\ (2010) 031.
\bibitem{Capitani:2011fg} S. Capitani et al, PoS\ LATTICE2011\ (2011) 145.
\bibitem{Bedaque:2004bx} P.~F.~Bedaque, Phys.\ Lett.\ B {\bf 593} (2004) 82.


\end{thebibliography}
\end{document}